\documentclass [aps,prb,preprint] {revtex4}
\usepackage{latexsym, epsfig, graphics}
\begin{document}

\preprint{LBNL-57632}

\title{How an antenna launches its input power into radiation: the pattern of the Poynting vector at and near an antenna}
\author{J. D Jackson\thanks{Electronic  mail: jdjackson@lbl.gov}}
\affiliation{Physics Department, University of California, Berkeley \\ and Lawrence Berkeley National Laboratory, Berkeley, California 94720}
\date{6 June 2005}

\begin{abstract}
In this paper I first address the question of whether the seat of the power radiated by an antenna made of conducting members is distributed over the ``arms'' of the antenna according to $ - {\bf J \cdot E}$, where ${\bf J}$ is the specified current density and ${\bf E}$ is the electric field produced by that source. Poynting's theorem permits only a global identification of the total input power, usually from a localized generator, with the total power radiated to infinity, not a local correspondence of $- {\bf J \cdot E}\  d^{3}x $ with some specific radiated power, $r^{2} {\bf S \cdot \hat{r}}\  d\Omega $. I then describe a model antenna consisting of two perfectly conducting hemispheres of radius \emph{a} separated by a small equatorial gap across which occurs the driving oscillatory electric field. The fields and surface current are determined by solution of the boundary value problem. In contrast to the first approach (not a boundary value problem), the tangential electric field vanishes on the metallic surface.  There is no radial Poynting vector at the surface. Numerical examples are shown to illustrate how the energy flows from the input region of the gap and is guided near the antenna by its ``arms'' until it is launched at larger \emph{r/a} into the radiation pattern determined by the value of \emph{ka}.
\end{abstract}
\normalsize
\pacs{????}
\maketitle

{\bfseries I. INTRODUCTION} \\

This paper is a didactic discussion of how the electromagnetic energy radiated by an antenna emerges from a localized source, is guided by the antenna's conductors, and ultimately shakes free to form the radiation described by the asymptotic Poynting vector. How this happens and what is the true seat of the power radiated is understood by many, but not by all. I hope that an analysis of the rights and wrongs and the treatment of specific examples prove useful.
  
The question of how electromagnetic energy is transported along a system of conductors dates back to J. H. Poynting's 1884 paper, \emph{On the transfer of energy in the electromagnetic field}, in which he enunciates his theorem and discusses various examples.~\cite{poynting}  Poynting does not consider radiating systems, but he was very clear on the transport of the electromagnetic energy associated with current-carrying wires and quasi-static circuits such as a discharging capacitor.

In modern notation Poynting's theorem takes the form,
\begin{equation}
- \ \int_{V}{\bf J \cdot E}\ d^{3}x = \int_{V}\frac{\partial u}{\partial t}\ d^{3}x\ + \ \oint_{S}{\bf n \cdot S}\ da \label{theorem}
\end{equation}
Here ${\bf S = E \times H}$ is the \emph{Poynting vector}, ${\bf E}$ and ${\bf H}$ are the fields, ${\bf J}$ is the current density, and, with some qualifying caveats that need not concern us here, $u = \frac{1}{2}(\bf{E \cdot D + B \cdot H})$ is the electromagnetic energy density within a chosen volume \emph{V} bounded by the surface \emph{S}. The physical interpretation of the theorem is that the left-hand integral represents the rate at which the given sources ($\bf{J}$) supply energy to the electromagnetic fields.  The volume integral on the right is the rate of increase of electromagnetic energy within \emph{V}, while the surface integral is the rate at which energy escapes from \emph{V} through the surface \emph{S}.  The theorem is basically a statement of conservation of energy.

In the late 19th and early 20th centuries, radiated systems were investigated chiefly by specification of simple oscillating current and charge distributions.  The asymptotic fields were found and the Poynting vector evaluated to give the radiation pattern.  Performing the surface integral on the right in (\ref{theorem}) gave the total radiated power \emph{P} in terms of the parameters of the source, e.g., its dipole moment. For an antenna with a given input current \emph{I}, the result was often expressed in terms of a \emph{radiation resistance} $R_{rad}$ through the relation, $P = I^{2}R_{rad}$.

In 1922, Brillouin~\cite{brill} pointed out an alternative method for calculating $R_{rad}$.  For steady-state sinusoidal oscillation of the source, the volume integral on the right in (\ref{theorem}) vanishes for a time average; the input power is equal to the power radiated. The radiation resistance can be computed equally well by evaluating the integral over the sources on the left-hand side of (\ref{theorem}).  This approach was taken up by many~\cite{pist, bech} because of its appeal: the current of each element along the arms of an antenna was apparently the real source of the radiated energy.

At that time (and to some extent today) many believed that energy was transported within the conductors carrying current.  Poynting's statements to the contrary had been forgotten.  The focus on specification of distributed current sources is widespread in many applications.  For antennas the approach works, more or less. Often antennas have thin straight conducting elements. If such elements have infinitesimal cross sections, it can be shown that an oscillating current is distributed sinusoidally along its length. The assumption of sinusoidal behavior proved to yield reasonably accurate results for angular distributions of radiation from antenna arrays. Specifying a plausible current source ${\bf J}$ became standard practice. In such discussions no attention is paid to the fact that a radiating antenna is an electromagnetic boundary value problem in which the current distributed on the arms of the antenna emerges as part of the solution, not as input. Rather, an ethereal current distribution is postulated and the resulting fields calculated.  This approach may be appropriate for electrons orbiting in a synchrotron or undergoing a quantum-mechanical transition in an atom, but not for a careful treatment of a realistic antenna with conducting surfaces.

The antenna as a boundary value problem has, of course, an honorable history beginning in 1897 with Pocklington~\cite{pock} who discussed the lowest mode of a perfectly conducting ring of wire (damped by radiation loss). Lord Rayleigh extended the work to the higher modes in 1912, with special attention to the damping.~\cite{rayleigh}   The history is summarized by R. W. P. King~\cite{king1} in his Introduction. Noteworthy are the works of L. V. King~\cite{lvk} and  Hall\'{e}n~\cite{hallen} in the 1930s on a long cylindrical antenna with a solution by means of an integral equation. In the early 1940s, Schelkunoff, in a paper~\cite{schel1} and a book~\cite{schel2}, discussed perfectly conducting biconical antennas in which the cones act as wave guides. A spherical antenna with a planar circular gap can be viewed as a limiting case.~\cite{schel2}  At the same time, Stratton and Chu analyzed three models of antennas: the circular cylinder~\cite{stratton1},the  sphere~\cite{stratton2}, and the prolate spheroid.~\cite{stratton3}  Subsequently, R. W. P. King and colleagues made extensive use of Hall\'{e}n's approach for linear antennas.  Their results and literature are thoroughly documented in King's book.~\cite{king1}  A sampling of the accumulating literature on cylindrical, biconical, and spherical antennas in the applied journals is contained in references.~\cite{silver,irmer,chang,hamid}

In none of these works is attention paid to the fields and energy flow in the immediate neighborhood of the antenna.  F. M. Landstorfer and co-workers are one group that has explored numerically field lines and power flow in the neighborhood of guiding conducting surfaces: a half-dipole antenna on a conducting plane,~\cite{lands1}  wave-guide discontinuities and dipole antennas,~\cite{lands2} two-dimensional diffraction,~\cite{lands3} and examples of gain-optimized antennas.~\cite{lands4}  Surprising eddies of power flow can be seen in some examples.

In Section II, I discuss a thin center-fed linear antenna by specification of a sinusoidal axial current density of vanishing cross section, with no attempt to solve a boundary value problem. The two different approaches to determining the total power radiated are described.  The question of whether the element of ``input'' power  $-{\bf J \cdot E}$ at a particular point along the antenna can associated with a particular segment of the radiated power is answered. I also make some comments here on reasons for the relative success of this na\"{i}ve approach. In Section III, I describe a particular antenna boundary value problem, a perfectly conducting sphere with a small gap at the equator, across which is an azimuthally uniform electric field.
As already stated, this problem has been addressed by Stratton and Chu~\cite{stratton2} and by Schelkunoff~\cite{schel3}. These authors were interested in the impedance of the antenna, including radiation, but did not discuss the Poynting vector in the neighborhood of the antenna or its evolution to the far fields. Section III sets up the formalism; Section IV gives examples of the energy flow, the current on the surface, and the modification of the energy flow for non-vanishing surface resistivity.  Section V contains concluding remarks.  An appendix gives the expansion in associated Legendre functions of the electric field at the gap, a necessary ingredient for the multipole expansion of Section II.\\

{\bfseries II. EXAMPLE OF LINEAR ANTENNA} \\

As discussed in the Introduction, a common approach to the emission of radiation by an antenna is to postulate 
(with greater or lesser justification) the sources as given current and 
charge distributions.  The potentials and the fields are then evaluated in 
the usual way as integrals over the source with the appropriate Green 
function.  We discuss briefly the very thin center-fed linear antenna in 
order to illustrate the confusion that can occur in evaluating the radiated 
power in different ways via Poynting's theorem and in arguing the source of the 
power.

The time dependence is assumed to be $exp(-i\omega  t)$ with $k = \omega/c$.  
Complex notation is used, with physical quantities as the real parts of 
complex expressions. The antenna is of length \emph{2a}, located on the 
\emph{z}-axis on the interval $-a<z<a$. The current and charge densities on 
that interval are assumed sinusoidal and are
\[\mathbf{J}(x, y,z) = \hat{z}\ I_{0}\ \ \sin(ka-k|z|)\ \delta(x)\delta(y) \]
\[\rho(x, y, z) = i \frac{I_{0}}{c}\ \epsilon(z)\ \ \cos(ka-k|z|) \ 
\delta(x)\delta(y) \]
\noindent With the time dependence suppressed, the vector and scalar potentials in the Lorenz gauge at a point \emph{(x, y, z)} are
\[\mathbf{A} = \hat{z}\frac{\mu_{0}I_{0}}{4\pi}\int_{-a}^{a} dz' \ \ \sin(ka-
k|z' |)\ e^{ikR}/R \]
\[\Phi = i\frac{Z_{0}I_{0}}{4\pi}\int_{-a}^{a} dz' \ \epsilon(z')\ \cos(ka-k|z' 
|)\ e^{ikR}/R \]
\noindent Here $Z_{0}=\sqrt{\mu_{0}/\epsilon_{0}}$  is the impedance of free 
space, $R = \sqrt{\rho^{2}+(z-z')^{2}}$,  $\rho = \sqrt{x^{2} + y^{2}}$, and $\epsilon(x) = \pm 1$ 
for $ x > 0, x< 0 $.  To evaluate the scalar potential's contribution to the 
electric field we need 
\[-\mbox{\boldmath $\nabla$}(e^{ikR}/R) = \mbox{\boldmath$\nabla' 
$}(e^{ikR}/R) \]
The electric field is the sum of the negative time derivative of the vector 
potential (${\bf E}_{1}$) and the negative gradient of the scalar potential  
(${\bf E}_{2}$) :
\begin{eqnarray} 
{\bf E}_{1} & = & i\frac{Z_{0}I_{0}}{4\pi}\int_{-a}^{a} dz' [\hat{z}\ k\ \sin(ka-
k|z'|)e^{ikR}/R  \\ 
{\bf E}_{2} & = & i\frac{Z_{0}I_{0}}{4\pi}\int_{-a}^{a} dz'\  
\epsilon(z')\ \cos(ka-k|z'|) \mbox{\boldmath $\nabla'$}(e^{ikR}/R)    
\end{eqnarray}
Taking only the \emph{z}-component of the gradient, we integrate by parts in 
${\bf E}_{2}$ to obtain
\begin{equation}
 E_{2z} = i\frac{Z_{0}I_{0}}{4\pi}\left[\ (e^{ikr_{1}}/r_{1}+ 
e^{ikr_{2}}/r_{2}) - \int_{-
a}^{a}dz'\frac{e^{ikR}}{R}\frac{\partial}{\partial z'}[\epsilon(z')\ \cos(ka-
k|z'|)]\:\right] 
\end{equation}
Here $r_{1}= \sqrt{\rho^{2}+(z-a)^{2}}$ and $ r_{2}= 
\sqrt{\rho^{2}+(z+a)^{2}}$.  The derivative of the charge density has a 
delta function contribution at the origin from the derivative of $\epsilon 
(z')$ so that, when we add in the transverse contribution we have
\begin{eqnarray}
{\bf E}_{2}& = &- {\bf E}_{1}  \nonumber \\
		 &   & \: +\  i\ \frac{Z_{0}I_{0}}{4\pi}\left[\ 
\hat{z}\left(e^{ikr_{1}}/r_{1}+ e^{ikr_{2}}/r_{2} - 2\ \cos(ka)e^{ikr}/r\right) 
\right. \nonumber\\
		 &   & \: +\int_{-a}^{a}dz'\ \left.\epsilon(z')\ \cos(ka-
k|z'|)\mbox{\boldmath $\nabla'_{\bot}$} (e^{ikR}/R)\: \right]
\end{eqnarray}

Here $r=\sqrt{\rho^{2}+z^{2}}$ is the distance from the origin.  The total 
electric field is therefore
\begin{eqnarray}
  {\bf E}& = & \ i\ \frac{Z_{0}I_{0}}{4\pi}\left[\ 
\hat{z}\left(e^{ikr_{1}}/r_{1}+ e^{ikr_{2}}/r_{2} - 2\ \cos(ka)e^{ikr}/r\right) 
\right. \nonumber \\
         &   & \: + \int_{-a}^{a}dz'\ \left. \epsilon(z')\ \cos(ka-
k|z'|)\mbox{\boldmath $\nabla'_{\bot}$} (e^{ikR}/R) \:\right] 
\end{eqnarray}

If we wish to evaluate the time-averaged power input $P_{1}$ to the electromagnetic field,
\begin{equation} 
P_{1} = -\ \int d^{3}x\ \langle{\bf J\cdot E}\rangle 
\end{equation} 
we need only the \emph{z} component of the electric field on the axis.
Explicitly, we have
\begin{equation}
P_{1}= -\frac{1}{2}Re\ \int_{-a}^{a}dz\ I_{z}E^{*}_{z}(\rho=0,z) 
\end{equation}
With 
\begin{eqnarray}
E_{z}(0,z) &=& i\ \frac{Z_{0}I_{0}}{4\pi} \left [e^{ik|a-z|}/|a-z| \right. \nonumber \\
		 & & \mbox{}+ \left. e^{ik|a+z|}/|a+z| -2\ \cos(ka)e^{ik|z|}/|z| \right 
] 
\end{eqnarray}
we find
\begin{eqnarray}
P_{1}&=&\frac{Z_{0}I_{0}^{2}}{8\pi}\int_{-a}^{a}dz \ \ \sin(ka-k|z|)   \nonumber	 \\
     &  &\times\frac{2a}{a^{2}-z^{2}}\left [\ \ \sin(ka)\ \cos(kz)-
(a/z)\ \cos(ka)\ \sin(kz)\ \right ] \label{p-1} 
\end{eqnarray}
The first line in the integrand is recognizable as the current; the second is 
the electric field after the use of some trigonometric identities.\\

By Poynting's theorem the input power $P_{1}$ is equal to the integral of the 
outward normal component of the time-averaged Poynting's vector ${\bf S}= 
Re({\bf E \times H^{*}})/2$ through any closed surface surrounding the 
antenna, in particular a sphere of large radius centered on the antenna. The 
well-known result~\cite{king2,jackson} for the Poynting-vector power $P_{2}$ is
\begin{equation}
P_{2}=\frac{Z_{0}I_{0}^{2}}{4\pi}\int_{-1}^{1}d(\ \cos\theta)\ \frac{|\ \cos(ka\ 
\ \cos\theta)-\ \cos(ka)|^{2}}{1-\ \cos^{2}\theta}  \label{p-2}
\end{equation}
We write with malice aforethought $\zeta=a\ \ \cos\theta$ and transform $P_{2}$ 
Into
\begin{equation}
P_{2}=\frac{Z_{0}I_{0}^{2}a}{4\pi}\int_{-a}^{a}d\zeta\ \frac{|\ \cos(k\zeta)-
\ \cos(ka)|^{2}}{a^{2}-\zeta^{2} } \ 
\end{equation}

Comparison of $P_{1}$ and $P_{2}$ suggests several questions:\\
1.  Are the two expressions actually equal?  The answer is yes. Each can be 
transformed into the same uninformative sum of constants, logarithms, and sine 
and cosine integrals~\cite{king2}  that undulates as a function of \emph{ka}.~\cite{king3,jack2} \\
\noindent 2.  Does the variable \emph{z} in $P_{1}$ correspond in any way to 
$\zeta=a\ \cos\theta$ in $P_{2}$ and so connect the increment of power at 
position \emph{z} on the antenna to the increment of radiated power at angle 
$\theta$ ?  The answer is no for several reasons. First of all, the integrand 
in $P_{1}$ is a spatial scalar, without direction, whereas the integrand in 
$P_{2}$, although a scalar, is the dot product of the Poynting vector with a 
unit radial vector whose direction changes with angle. 

\begin{figure}[htp]
\centering
\includegraphics[width=4in]{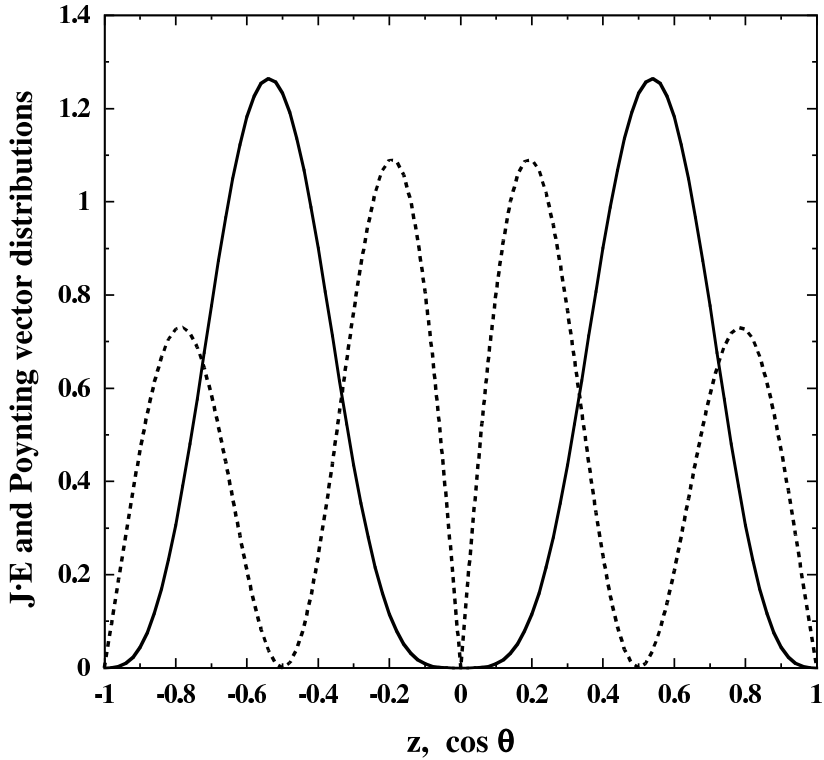}
\vspace{.25in}
\caption{\small Normalized integrands for the two forms $P_{1}$ and $P_{2}$ for the total
 radiated power from a linear antenna with $ka=2\pi$. The solid curve is the angular distribution of radiation in $\ \cos\theta$ from the Poynting vector expression $P_{2}$. The dotted curve is the integrand ${\bf J \cdot E}$ of $P_{1}$ versus \emph{z/a}.  }
\label{linear}
\end{figure}

Secondly, a comparison of integrands shows smaller and larger differences between the two ``angular'' distributions depending on the value of \emph{ka}. As a matter of fact, for $ka= (2n+1)\pi/2, \ \ n=0,\ 1,\ 2, \ldots \ $, the integrand of (\ref{p-1}) as a function of $z/a$ is equal to that of (\ref{p-2}) for $\ \cos\theta = z/a$, but for other values of \emph{ka} the integrands can be vastly different, as shown for example in Figure~\ref{linear} for $ka = 2\pi$.
\\
\normalsize
\noindent 3.  Can the expression for $P_{1}$ justify the claim that the power radiated 
by an antenna has its origins at various points along the length of the 
antenna? The short answer is no.

First of all, the fields at any point in space receive coherent contributions from 
all elements of the current along the whole antenna. King~\cite[p.562-5]{king1} 
discusses a center-fed antenna with $ka = 3\pi/2$, imagining it as the limit of a 
line of closely spaced individual generators that produce the sinusoidal current 
distribution. He shows that \emph{in a certain sense} one can identify each of the three 
separate ``ears'' of the angular distribution at infinity with a corresponding 
half-wavelength segment of the current distribution.  But he is at great pains to 
make clear it \emph{does not mean} that the fields in each separate angular region 
are ``maintained entirely and exclusively by the current in [the corresponding] 
third of the antenna. \ldots The field at any point in space is maintained by 
\emph{all} currents in the entire antenna.'' (\emph{op. cit.}, p.565).

Secondly, as stressed in the Introduction, a source distribution $ ({\bf J},\rho)$ in the Maxwell equations does 
not define appropriately an antenna boundary-value problem. Antennas are almost 
universally constructed of conductors, usually very good conductors. An excellent approximation for the boundary-value problem is to assume perfect conductivity. Then there can be a 
surface current density, but \emph{no tangential electric field at the 
surface}. There is therefore no $\bf J\cdot E$ to integrate over the surface 
to find the power. You might say, well, in the real world conductivities are 
not infinite. Suppose we assume excellent, but not perfect, conductivity. In 
that situation things go the opposite way: There \emph{is} now a tangential 
electric field at the surface, but \emph{the Poynting vector points inward!} 
Energy flows inward into, not out from, the conductor; the resistance of the 
metal converts the electromagnetic energy into heat; the antenna robs power 
from the source~\cite{poynting,sommerfeld}.\\ 

As already said in the Introduction, most antenna experts  seem to accept the idea that the antenna structure does not itself radiate (in the sense that segments of the conductor are the local sources of power), but rather guides the energy from the input source and launches it into its final radiation pattern. Why then is the approach of specification from the beginning of the current on an antenna so widely accepted and employed?  One reason is that the most common antenna is linear, or an array of linear elements.  For a thin linear antenna the zeroth order current is sinusoidal in amplitude and constant in phase with position. The solution~\cite{hallen} of the boundary value problem yields corrections to the current as a series in powers of $\Omega^{-1}$, where $\Omega = 2ln(2h/r)$, with $2h$ the length of the antenna and \emph{r} its radius. Values of $\Omega \approx 10 \pm 3$ is a plausible range: a \emph{100 MHz} $\lambda/2$ antenna with $2h = 1.5 \ m$ and a diameter of \emph{0.5 (1.0) cm} has $\Omega = 13 (10)$.

Calculations including first order corrections show that for short antennas ($kh \leq 2$) the real part of the current either dominates the imaginary part over almost the whole length of the antenna, or the relative phase changes little over the length. In either circumstance the radiation pattern of an antenna or antenna array will differ little from that computed with the na\"{i}ve zeroth order current.  Even for larger \emph{kh}, for which the zeroth order current changes sign along the length, the imaginary part of the current is generally small compared to the real part (except in the narrow regions where the real part vanishes).  Examples of the current components on center-fed linear antennas of different \emph{kh} are given by Hall\'{e}n~\cite{hallencu} and King.~\cite{kingcu}  

If reliable values of input impedance are desired, second order corrections to the current at the input point may be required.~\cite{kingsecond}  But for \emph{radiation patterns} of linear antennas the na\"{i}ve assumption of a real sinusoidal current is adequate in most circumstances. 

We now turn to a solvable model of an antenna to show in detail how the energy flows in its neighborhood and beyond.\\

{\bfseries III. CENTER-FED SPHERICAL ANTENNA} \\

To illustrate the near fields and power flow in the immediate neighborhood of an antenna, we consider a spherical antenna of radius \emph{a} with excitation provided by an electric field across a narrow equatorial gap between the perfectly conducting near-hemispheres.  As noted earlier, this problem has been treated in some respects by Stratton and Chu~\cite{stratton2} and Schelkunoff.~\cite{schel3}  We follow the notation of Section 9.7 ff in my text.~\cite{jackson} 
The power source does not concern us except that it manifests itself as an electric field across the gap (and the accompanying surface current on the surface of the sphere). The source can be imagined in Schelkunoff's language of conical antennas. Our antenna can be thought of as a limiting case of a biconical antenna with spherical segments as caps and an opening angle of almost \emph{90} degrees. The sides of the cones meet at the origin inside the sphere. It is there that an infinitesimal generator resides. It causes an electric field between the upper and lower cones to emerge in an azimuthally symmetric way as $E_{\theta}$ across the gap.\\

{\bfseries A. Fields and Surface Current}\\

Outside the sphere the electric and magnetic fields can be described by multipole fields.~\cite{jack3}  The magnetic multipole fields have a non-vanishing radial component of the magnetic field, while the electric multipole fields have a non-vanishing radial component of the electric field. Because the normal component of magnetic field vanishes at the surface of a perfect conductor, while the normal component of the electric field does not, we conclude that the spherical antenna has \emph{only electric multipole fields}. Furthermore, because of the assumed azimuthal symmetry, only $(\ell,m)$ multipoles with $m=0$ occur. 
The electric and magnetic fields for $r\geq a$ that represent outgoing waves at infinity are
\begin{eqnarray}
\bf{E} & = & -Z_{0}\sum_{\ell}\ \frac{a(\ell)}{\sqrt{\ell(\ell+1)}}\left[\mbox{\boldmath    $\hat{\theta}$}\ D_{\ell}(kr)\frac{\partial}{\partial\theta}Y_{\ell,0}\ + \ \mbox{\boldmath$\hat{ r}$}\ \ell(\ell+1)\frac{h_{\ell}^{(1)}(kr)}{kr}Y_{\ell,0}\right]   \label{bfE}\\
\bf{H} &= & -i\ \mbox{\boldmath$\hat{ \phi}$}\  \sum_{\ell}\frac{a(\ell)}{\sqrt{\ell(\ell+1)}}h_{\ell}^{(1)}(kr)\ \frac{\partial}{\partial\theta}Y_{\ell,0} \label{bfH}
\end{eqnarray}
Here $k=\omega/c$, $a(\ell)$ is the $\ell^{th}$ electric multipole amplitude, $h_{\ell}^{(1)}$ is the spherical Hankel function of the first kind, and the function $D_{\ell}(x)$ is
\begin{equation}
D_{\ell}(x) = \frac{1}{x}\frac{d}{dx}[x\ h_{\ell}^{(1)}(x)]\ =\ \left (\frac{h_{\ell}^{(1)}(x)}{x}\ +\ \frac{dh_{\ell}^{(1)}(x)}{dx}\right)
\label{D(x)}
\end{equation}
With $\partial Y_{\ell,0}/\partial \theta = \sqrt{(2\ell+1)/4\pi}\ P_{\ell}^{1}(\ \cos\theta)$, the $\theta$ component of the electric field can be written as
\begin{equation}
E_{\theta}= -\ Z_{0}\sum_{\ell}\sqrt{\frac{2\ell+1}{4\pi\ell(\ell+1)}}\ a(\ell)\ D_{\ell}(kr)\  P_{\ell}^{1}(\ \cos\theta) \label{E-theta}
\end{equation}
To determine the multipole coefficients $a_{\ell}$ we must equate this component of the field at $r=a$ to the expansion found in Appendix A for the field across a symmetric gap defined by angles $\pm \epsilon$ on either side of $\theta = \pi/2$. Equating (\ref{E-theta}) with (\ref{A7}) yields
\begin{equation}
a(\ell)D_{\ell}(ka)\ =\ \frac{V}{Z_{0}a}\sqrt{\frac{\pi(2\ell+1)}{\ell(\ell+1)}}\ \frac{P_{\ell}(\ \sin\epsilon)}{\epsilon}\ \  \; \; (\ell \  odd) \label{a-ell}
\end{equation}
For a symmetric equatorial gap, the even $\ell$ multipole moments vanish.
The surface current on the hemi-spherical conductors is given by $\mbox{\boldmath$K=\hat{r}\times  H$}(r=a+)$.  There is only an azimuthal component of magnetic field (\ref{bfH}).  The surface current is therefore in the $\theta$ direction:
\begin{equation}
K_{\theta}= -H_{\phi}(r=a+)\ = \ i\ \sum_{\ell\ odd}\sqrt{\frac{(2\ell+1)}{4\pi\ell(\ell+1)}}\ a(\ell)\ h_{\ell}^{(1)}(ka)\ P_{\ell}^{1}(\ \cos\theta) \label{K-theta}
\end{equation}\\

{\bfseries B. Total Power Input from the Source}\\
 
The oscillating electric field and the associated magnetic field in the gap produce a radial power flow at $r = a$. This time-averaged input power is given by the integral over the segment of the sphere occupied by the gap of $Re({\bf E \times H^{*}})/2$:
\begin{equation}
P_{\rm input}\ = \ \frac{1}{2}Re \left (a^{2}\int_{0}^{2\pi}d\phi\ \int_{-\sin\epsilon}^{\ \sin\epsilon} d(\ \cos\theta)\;E_{\theta}\ H_{\phi}^{*} \right ) 
\end{equation}
In Appendix A it is shown that the field $E_{\theta}$ in the gap region $(-\ \sin\epsilon  < \ \cos\theta < \ \sin\epsilon)$ is 
\[E_{\theta}(r=a)\ = \frac{V}{2a\epsilon\ \ \sin\theta} \]
With $H_{\phi}^{*}$ taken from (\ref{K-theta}), the input power is therefore
\begin{equation}
P_{\rm input}\ =  \pi a^{2}\int_{-\ \sin\epsilon}^{\ \sin\epsilon}\ d(\ \cos\theta)\ \ Re \left (\ \frac{i \ V}{2a\epsilon \ \sin\theta}\ \sum_{\ell \ odd}\sqrt{\frac{(2\ell+1)}{4\pi\ell(\ell+1)}}\ a^{*}(\ell)\ H_{\ell}^{(2)}(ka)\ P_{\ell}^{1}(\ \cos\theta) \right )  \label{P-in-1}
\end{equation}
With the definition of the associated Legendre function, \[P_{\ell}^{1}(\ \cos\theta)=-\ \sin\theta \ \partial P_{\ell}(\ \cos\theta)/\partial( \ \cos\theta) \] 
the integral is elementary.  The result for the input power is
\begin{equation}
P_{\rm input}=\ Re \left (-i \frac{\pi Va}{\epsilon}\ \sum_{\ell \ odd}\ \sqrt{\frac{(2\ell+1)}{4\pi\ell(\ell+1)}}a^{*}(\ell)\ P_{\ell}(\ \sin\epsilon)\ h_{\ell}^{(2)}(ka) \right )   \label{P-in-2}
\end{equation}
Comparison of the terms in this sum with $a(\ell)$ given by (\ref{a-ell}) shows that
\begin{equation}
P_{\rm input}\ = \ \frac{Z_{0}a^{2}}{2}\ \sum_{\ell\ odd}\ |a(\ell)|^{2}\ Re[-iD_{\ell}(ka)\ h_{\ell}^{(2)}(ka)\ ]  
\end{equation}
Now the Wronskians of the spherical Bessel functions can be used to show that
\begin{equation}
-i\ D_{\ell}(x)\ h_{\ell}^{(2)}(x)\ = \ \frac{1}{x^{2}}\ -\ \frac{i}{2}(\frac{d}{dx}+\frac{2}{x})[j_{\ell}^{2}+n_{\ell}^{2}]  
\end{equation}
Therefore the time-averaged input power becomes the standard multipole expression,~\cite{jack4}
\begin{equation}
P_{\rm input}\ = \ \frac{Z_{0}}{2\ k^{2}}\ \sum_{\ell \ odd}\ |a(\ell)|^{2} \label{P-in-3}
\end{equation} 
Note, however, that this result for the power is found as the input power at the source, not from integration of $r^{2}$ times the asymptotic radial Poynting's vector over all angles. \\

{\bfseries IV. POYNTING VECTOR AND LOCAL ENERGY FLOW} \\

{\bfseries A. Poynting vector} \\

The fields around the spherical antenna are such that the Poynting vector has components in the radial and transverse ($\theta$) directions. In terms of the fields in Section III.A, the time-averaged Poynting vector is
\begin{eqnarray}
\bf{S} & = &\frac{Z_{0}}{2}\ \sum_{odd \ \ell,\ell'}\ Re\left[\ \mbox{\boldmath$\hat{r}$}\ R(\ell,\ell', r) \ P_{\ell'}^{1}(\ \cos\theta) P_{\ell}^{1}(\ \cos\theta)  \right. \nonumber\\ 
       &  &\; +\left.\mbox{\boldmath$\hat{\theta}$}\ T(\ell, \ell', r)\ P_{\ell'}(\ \cos\theta)P_{\ell}(\ \cos\theta) \right ] \label{poynting}
\end{eqnarray}
The coefficients $R(\ell, \ell', r)$ and $T(\ell, \ell', r)$ are
\begin{eqnarray}
R(\ell, \ell', r) & = & \frac{-i}{4\pi}\sqrt{\frac{(2\ell+1)(2\ell'+1)}{\ell\ell'(\ell+1)(\ell'+1)}}\ a^{*}(\ell')a(\ell)\ h_{\ell'}^{(2)}(kr)D_{\ell}(kr) \label{R0}\\
T(\ell, \ell', r) & = & -\ \frac{\ell(\ell+1)}{kr}\ \frac{h_{\ell}^{(1)}(kr)}{D_{\ell}(kr)}\ R(\ell, \ell', r) \label{T0}
\end{eqnarray}

Before discussing the flow of power near the antenna, we consider two limits of the real part of the coefficient $R(\ell, \ell', r)$, when $\ell' = \ell $ and $kr \gg \ell,\  \ell'$. Using (\ref{D(x)}) and the asymptotic forms of the spherical Hankel functions, we find
\begin{eqnarray}
Re[\ R(\ell, \ell, r)] & = & \frac{1}{4\pi k^{2}r^{2}}\frac{(2\ell+1)}{\ell(\ell+1)}\ |a(\ell)|^{2} \label{R1}\\
\lim_{kr \ \gg \ \ell,\ \ell'}\ Re[R(\ell, \ell', r)] & = & \ \frac{1}{4\pi k^{2}r^{2}}\sqrt{\frac{(2\ell+1)(2\ell'+1)}{\ell\ell'(\ell+1)(\ell'+1)}}\ Re[(i)^{\ell' - \ell}a^{*}(\ell')a(\ell)]  \label{R2}\\
\lim_{kr \ \gg \ \ell,\ \ell'}\ [T(\ell, \ell', r)] & = &\ \frac{i\ell(\ell+1)}{kr}\ R(\ell, \ell', r) \label{T1}
\end{eqnarray}
Note that (\ref{R1}) holds for all $r \geq a$. Thus if the radial part of (\ref{poynting}) is integrated over a sphere of radius \emph{r}, the orthogonality of the associated Legendre functions,
\[ \int_{-1}^{1}P_{\ell'}^{1}(\ \cos\theta)P_{\ell}^{1}(\ \cos\theta)\ d(\ \cos\theta) \ =\ \frac{2\ell(\ell+1)}{2\ell+1}\delta_{\ell',\ \ell} \]  \\
plus (\ref{R1}) leads directly to (\ref{P-in-3}), valid at any $r \geq a $. Obviously this must hold because of conservation of energy flow. 

The other limiting form (\ref{R2}) shows that the asymptotic angular distribution of radiation depends on the real part of  $(i)^{\ell'-\ell}a^{*}(\ell')a(\ell)$ , whereas the radial power flow at nearer distances can be expected to be different because of the more complicated structure of the full expression (\ref{R0}). And at close distances there is power flow in the transverse ($\theta$) direction as well. However, the asymptotic form of \emph{T} (\ref{T1}) shows that that component of the Poynting vector falls off faster than $r^{-2}$. We explore these aspects immediately.\\
\begin{figure}[!b]
\centering
\includegraphics[width=5in]{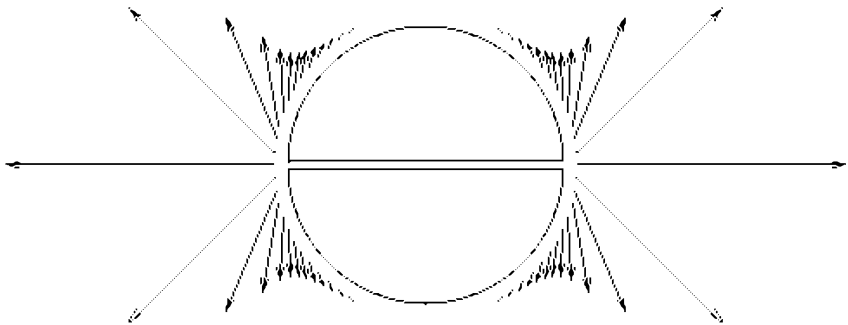}
\vspace{-0.5in}
\caption{\small Distribution of Poynting vector for $ka = 1.0$ and $r/a = 1.1$. Bases of arrows are observation points, lengths of arrows are relative magnitudes. Close to antenna, power squirts out from gap with some transverse flow guided along surface}
\end{figure}
\normalsize

{\bfseries B.  Examples of energy flow near the antenna} \\

Figure 2 shows the Poynting vector pattern close to the antenna for $ka = 1.0$. Here, and everywhere else unless stated otherwise, the total angular gap is $2\epsilon = \pi/50$. One sees that the power squirts out of the gap in a radial direction, with some transverse flow parallel to the surface, guided by the antenna surface.  Figure 3 is an alternative display, with $r/a = 10$ as well as $r/a=1.1$. The appreciable transverse flow and sharply peaked radial flow near the antenna goes over into almost purely radial flow of the asymptotic angular distribution.
\begin{figure}[htp]
\centering
\includegraphics[width=4in]{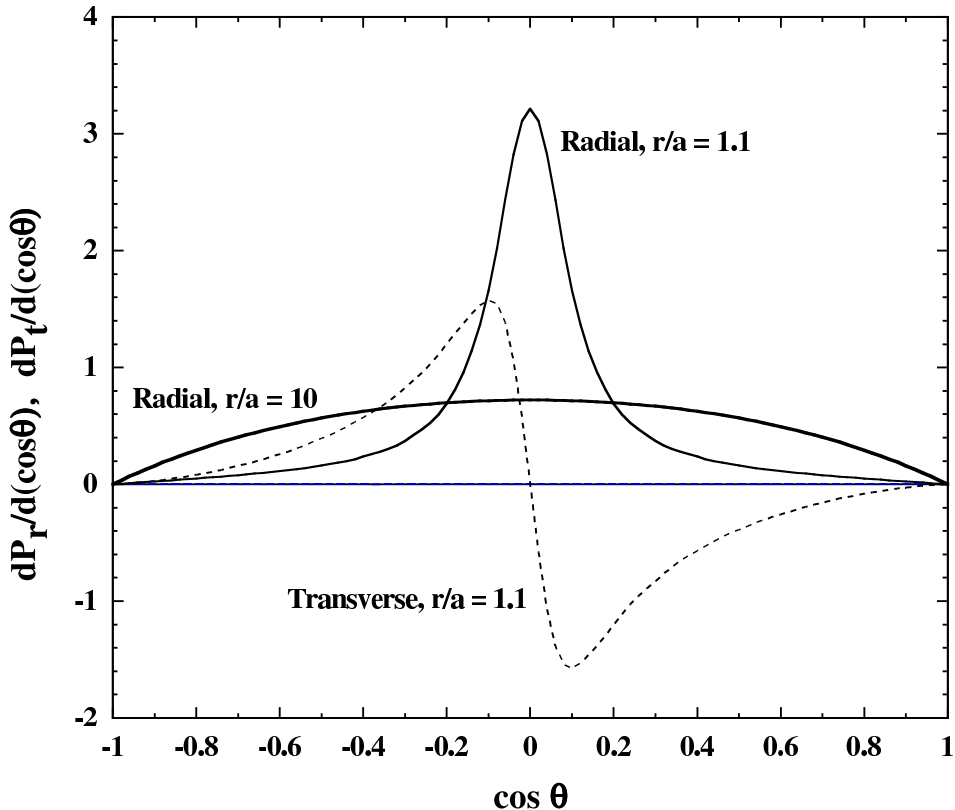}
\vspace{.5in}
\caption{\small Normalized distributions in $\ \cos\theta$ of radial and transverse power flow $dP_{r}/d(\ \cos\theta)$, $dP_{t}/d(\ \cos\theta)$  for $ka = 1.0$ and $r/a = 1.1,\  10$. Close to antenna, the power is peaked strongly in the equatorial plane, with some transverse flow. At $r/a = 10$, the angular distribution is essentially the featureless asymptotic form, with negligible transverse flow.}
\end{figure}
\normalsize
\indent Figures 4 and 5 show the corresponding features for $ka = 3\pi/2$ with $r/a = 1.1,\ 4$ in Figure 4 and $r/a = 1.1, \ 2, \ 10 $ in Figure 5. Note that the behavior of the power flow close to the antenna is essentially independent of the value of \emph{ka}, a consequence of the source being a small equatorial gap. The flow rearranges itself at moderate and large distances, however, into very different asymptotic forms for the different values of \emph{ka}. The antenna surface guides the flow only fairly nearby.\\
\begin{figure}[!ht]
\centering
\vspace{0.5in}
\includegraphics[width=4in]{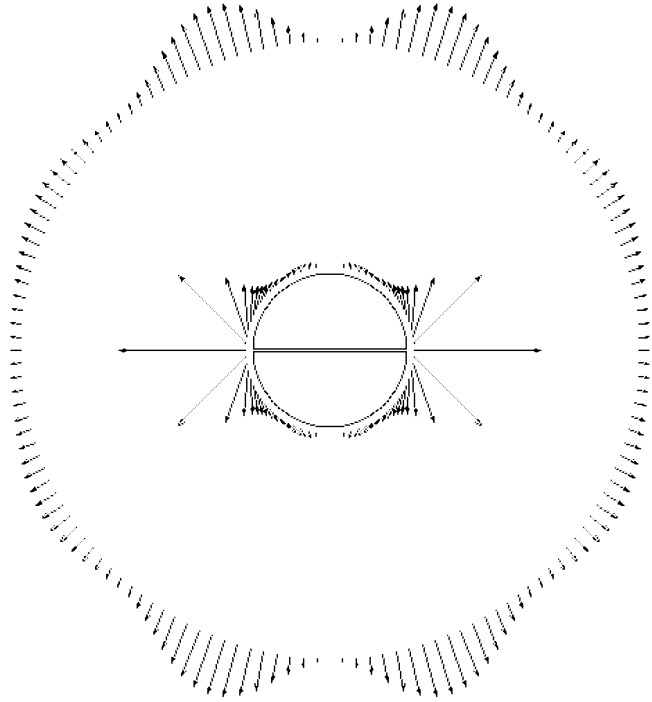}
\caption{\small Distribution of Poynting vector (times $r^{2}$) for $ka = 3\pi/2$ and $r/a = 1.1, 4.0$. Bases of arrows are observation points, lengths of arrows are relative magnitudes. Close to antenna, power squirts out from gap with some transverse flow along surface. At greater distance, flow is largely radial and begins to approach the asymptotic distribution.}
\end{figure}
\normalsize

\begin{figure}[htp]
\centering
\includegraphics[width=4in]{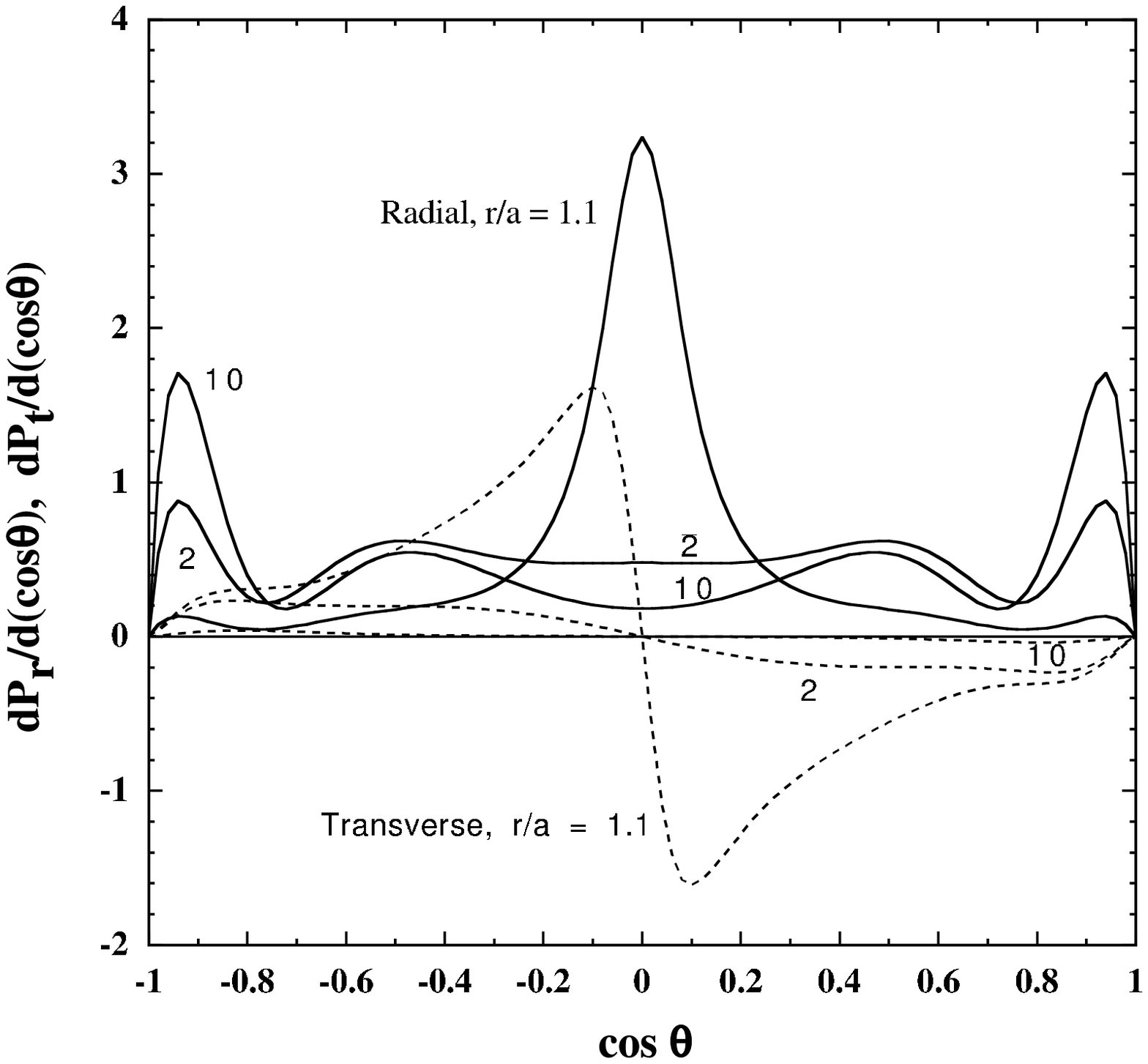}
\vspace{.3in}
\caption{\small Normalized distributions in $\ \cos\theta$ of radial (solid lines) and transverse (dashed lines) power flow $dP_{r}/d(\ \cos\theta), \ dP_{t}/d(\ \cos\theta)$  for $ka = 3\pi/2$ and $r/a = 1.1,\ 2, \   10$. Close to antenna, the radial power is peaked strongly in the equatorial plane, with considerable transverse flow. At $r/a = 2$, the equatorial peak has vanished, the transverse flow has diminished, and the radial flow begins to resemble the $r/a = 10$ angular distribution, which is essentially the asymptotic form, with negligible transverse flow.}
\end{figure}
\normalsize

{\bfseries C.  Examples of Surface Current} \\

The surface current density on the sphere is given by (\ref{K-theta}).  Integration over the surface in azimuth for fixed $\cos\theta$ yields the distribution $I_{theta} = 2\pi a \ \ \sin\theta \ K_{\theta}$, the total current flow across the ``latitude'' circle at fixed $\theta$, as the analog of the current $I(z)$ for a linear antenna. [Actually, $I_{sphere}(z) = \ \sin\theta \ I_{\theta}$ is a more accurate analog.]  The real and imaginary parts of $I_{\theta}$ in units of $V/Z_{0}$ are shown in Figure 6 for the example of $ka = 1.0$. The real part of the \emph{z}-component of the current is not far from proportional to the current on a linear antenna with $ka = 1$, namely $I_{0}\ \sin(1-|\ \cos\theta|)$.  Note, however, the comparable imaginary part, in contrast to the thin linear antenna.   
\begin{figure}[ht!]
\centering
\includegraphics[width=4in]{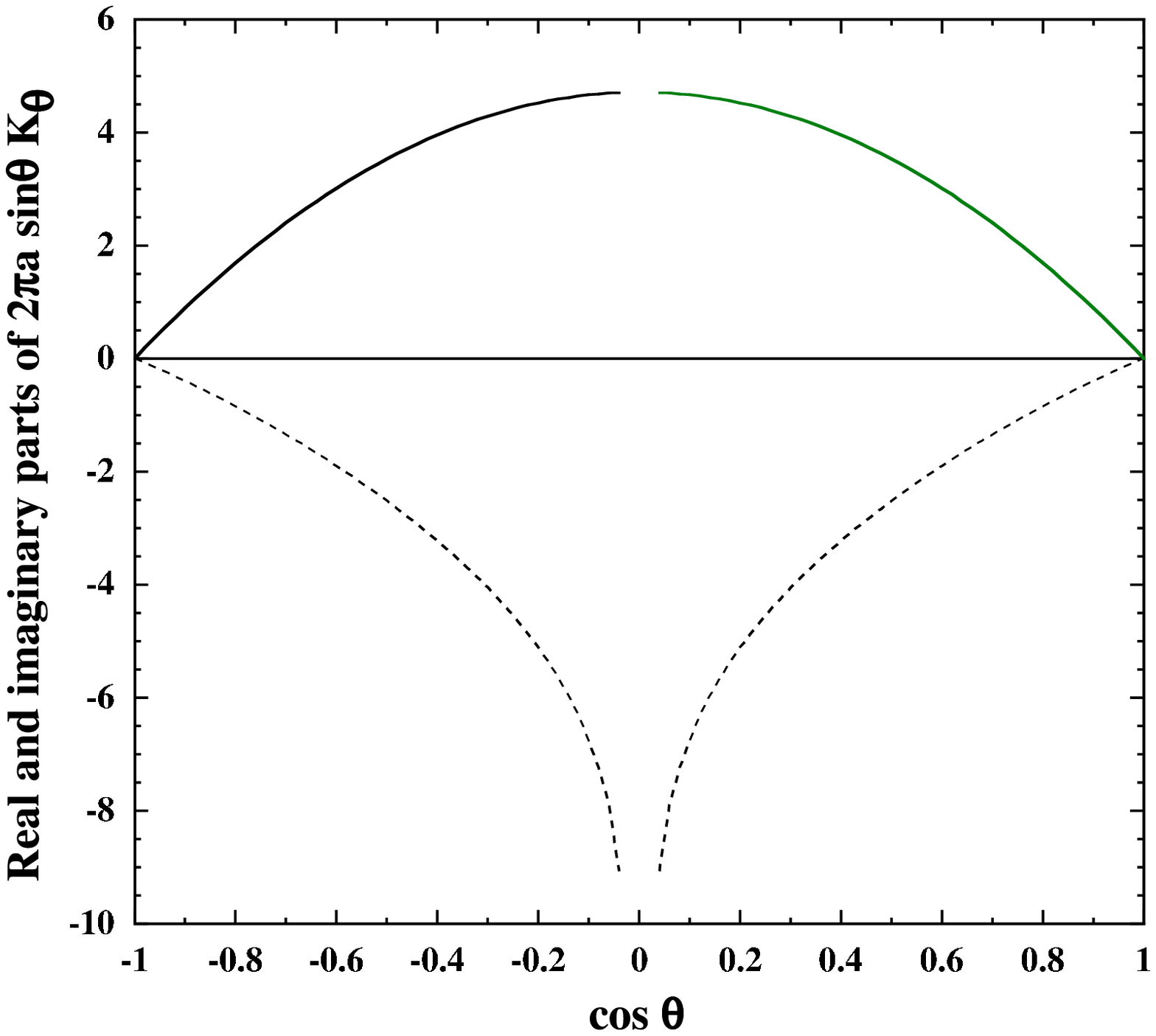}
\vspace{.2in}
\caption{\small Real (solid line) and imaginary (dashed line) parts of the total surface current $I_{\theta} = 2\pi a \ \ \sin\theta \ K_{\theta}(\ \cos\theta)$ in units of $V/Z_{0}$ versus $\ \cos\theta = z/a$ for an antenna with $ka = 1.0$.}
\end{figure}
\normalsize

The corresponding current for $ka = 3\pi/2$ is shown in Figure 7. Here the real part of the current displays some resemblance to the three half cycles of a sinusoidal current on a center-fed linear antenna of the same \emph{ka} despite the differences in shape.\\
\begin{figure}[ht!]
\centering
\includegraphics[width=4in]{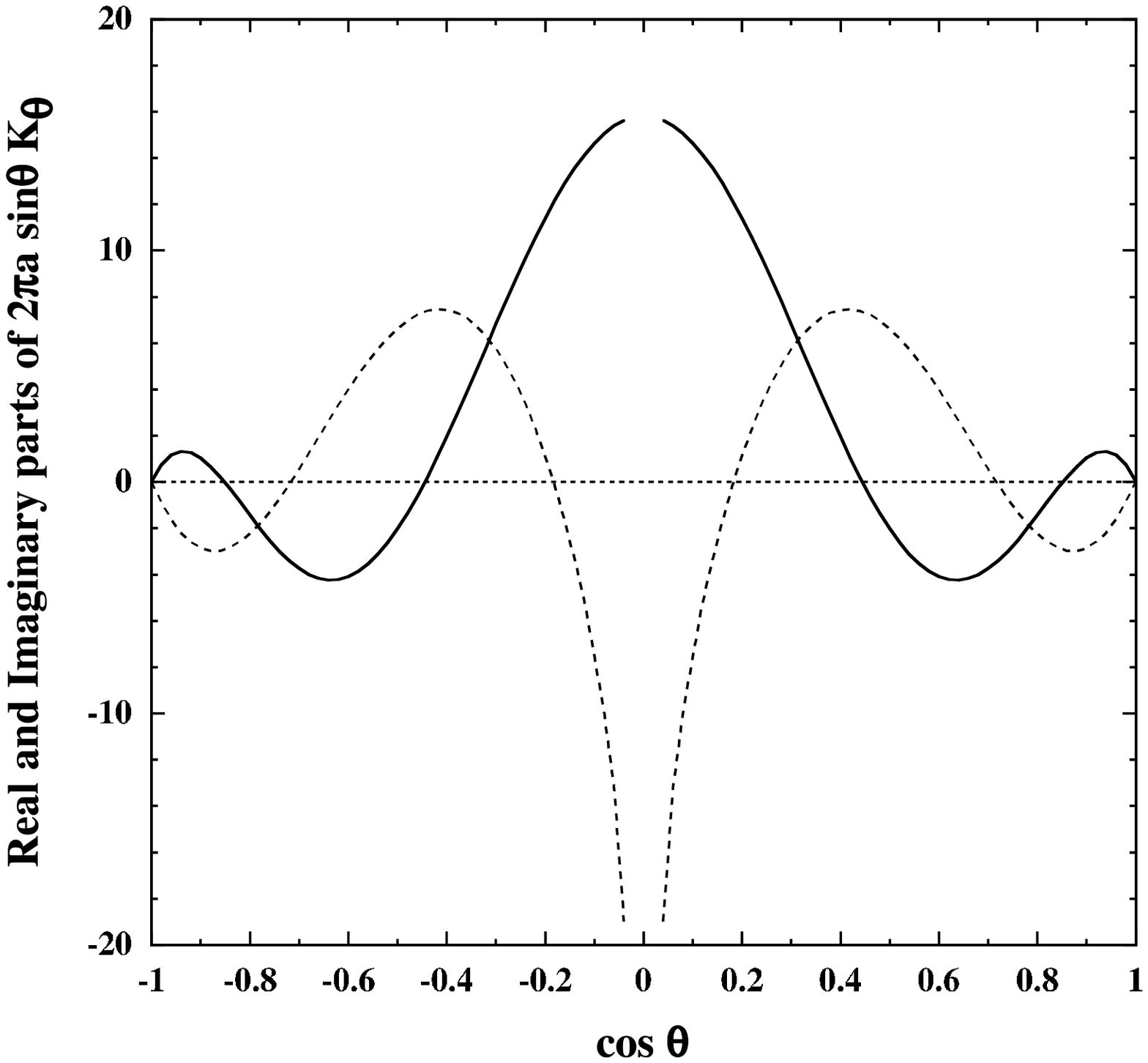}
\vspace{.2in}
\caption{\small Real (solid line) and imaginary (dashed line) parts of the total surface current $I_{\theta} = 2\pi a \ \ \sin\theta \ K_{\theta}(\ \cos\theta)$ in units of $V/Z_{0}$ versus $\ \cos\theta = z/a$ for an antenna with $ka = 3\pi/2$.}
\end{figure}
\normalsize

{\bfseries D.  Modifications because of finite conductivity}\\

The treatment so far has been based on vanishing resistivity on the surface of the antenna.  In such circumstances the only tangential electric field at $r=a$ is in the equatorial gap.  With small but non-vanishing resistivity, a small tangential electric field exists over the whole antenna. A perturbation approach can be used to find the fields and the \emph{inflowing} Poynting  vector at the surface.~\cite{jack5}  In addition to the zeroth order electric field (\ref{bfE}), there is a first order tangential electric field at the surface, given by
\begin{equation}
{\bf E^{(1)}}\ =\ \mbox{\boldmath $\hat{\theta}$}R_{s}(-1+i) H_{\phi}^{(0)}(r=a)
\end{equation}
where $R_{s}= \rho/\delta$ is the surface resistance, $\rho$ being the resistivity and $\delta$ being the skin depth. The magnetic field $H_{\phi}^{(0)}$ is given by (\ref{bfH}).  In passing we note that finite resistivity causes a modification in the multipole amplitudes (\ref{a-ell}):
\begin{equation}
a(\ell) \ \rightarrow \ a(\ell)\left[1\ -\ \frac{R_{s}}{Z_{0}}(1+i)\frac{h_{\ell}^{(1)}(ka)}{D_{\ell}(ka)} \right]
\end{equation}

The time-averaged energy dissipation per unit area of the antenna can be calculated either from the real part of the inward Poynting vector at the surface,

\[ -\mbox{\boldmath $\hat{r}$}\cdot \frac{1}{2} \ Re[{\bf E^{(1)}\times H^{(0)}}] \] 
or from one half the surface resistance times the square of the surface current density,
\begin{equation}
\frac{dP_{loss}}{dA}\ =\ \frac{1}{2}R_{s}|K_{\theta}|^{2}
\end{equation}
Here $|K_{\theta}| = |H_{\phi}^{(0)}|$, as shown in (\ref{K-theta}). Note that, in the Poynting vector expression, the zeroth order electric field does not appear; its contribution gives rise to the source-generated \emph{outward} power flow at the gap. For orientation, we note that for copper at room temperature, $R_{s}/Z_{0}=6.8 \times 10^{-7}, \ 2.2 \times 10^{-5}$, at $1 MHz, \ 1 GHz$, respectively. For aluminum alloys the numbers are 1.5 - 2.0 times larger.

An example of the distribution of energy loss over the surface of the antenna is given in Figure 8 for $ka = 3\pi/2$.  The dimensionless quantity displayed is the absolute square of the ratio of the surface current density at $\ \cos\theta$ to its value at $\ \cos\theta = \ \sin\epsilon$, the edge of the gap. This quantity is related to the power loss according to
\begin{equation}
\frac{1}{P_{rad}}\frac{dP_{loss}}{d(\ \cos\theta)}\ =\frac{R_{s}}{2\pi  R_{rad}}\ \frac{|K_{\theta}(\ \cos\theta)|^{2}}{|K_{\theta}(\ \sin\epsilon)|^{2}}    \label{loss} 
\end{equation}

\begin{figure}[htp]
\centering
\includegraphics[width=4in]{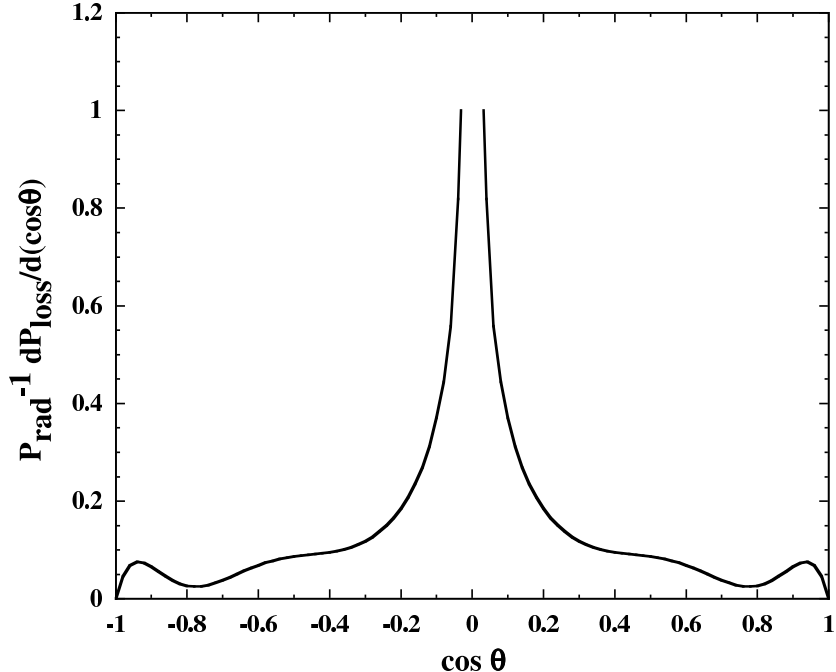}
\vspace{.2in}
\caption{\small Distribution of resistive power loss per unit $\ \cos\theta$ for $ka = 3\pi/2$. The plotted quantity is the absolute square of the ratio of surface current densities given in equation (\ref{loss}).  The losses are largest near the gap, as that is the region of largest current - see Figure 7.} 
\end{figure}
\normalsize

The ratio of the total dissipative loss to the radiated power is found by integrating (\ref{loss}) over $\ \cos\theta$:
\begin{equation}
\frac{P_{loss}}{P_{rad}}\ =\ \frac{R_{s}}{Z_{0}}\ \frac{\sum_{odd \ \ell}|a(\ell)|^{2}|ka \ h_{\ell}^{(1)}(ka)|^{2}}{\sum_{odd \ \ell}|a(\ell)|^{2}}
\end{equation}
The factors $|xh_{\ell}^{(1)}(x)|^{2}$ in the numerator are polynomials in inverse powers of $x^{2}$ up to $x^{-2\ell}$ (For $\ell=1$ the polynomial is ($1+x^{-2}$)).  For $ka << 1$, the dominant multipole is $\ell=1$; the ratio of the sums is closely $1/(ka)^{2}$.  An alternative way of looking at it is that, in the quasistatic limit, the dissipative losses are frequency independent (except for that in the surface resistance), while the power radiated is proportional to $(ka)^{2}$.

For the opposite limit, $ka >> \ell $, the spherical Hankel function factor can again be large compared to unity, causing successive terms in the numerator to fall off more slowly than those in the denominator.  But since the initial terms in numerator and denominator are comparable, the cumulative effect is not major and the ratio of sums is of order unity. Here are some examples: $(ka;\ ratio) = (0.1;\ 119.6),\ (0.5; \ 5.62),\ (1.0;\ 2.21),\ (10;\ 2.62),\ (20;\ 2.66)$. \\

{\bfseries V. CONCLUDING REMARKS}\\

The conventional way of specifying the plausible sources $\rho,\ {\bf J}$ of charge and current on antennas can yield reasonable radiation patterns, but fails to address the actual boundary value problems. By discussing first a conventional example of a center-fed linear antenna, I have, I hope, shown that the idea of a one-to-one correspondence between an increment of input power $ -{\bf J \cdot E} d^{3}x $ in or on the antenna and an increment of radiated power per unit solid angle $ r^{2}{\bf S \cdot \hat{r}}$ is without merit. Only the total input power can be equated to the total radiated power. Treatment of a tractable, albeit stylized, perfectly conducting spherical antenna with gap excitation as a boundary value problem illustrates how the antenna's current distribution emerges as part of the solution.  The fields right down to $r = a $ permit calculation of the Poynting vector everywhere outside the antenna.  Numerical examples show how the power flow originates at the gap, is guided near the antenna by the conducting surface of the antenna, and is launched toward its ultimate radiation pattern after a few multiples of \emph{r/a}.  With perfect conductivity there is no energy flow into or out from the antenna's surfaces.  When resistive losses are included, a small radial component of the Poynting vector occurs at the surface of the antenna, directed \emph{inward} into the conductor where it is dissipated in ohmic heating. 

The main message is that antennas are boundary value problems, that the ``arms'' of an antenna guide and launch the energy flow but are not its origin, and that initial specification of the current for a metallic antenna is not consistent (even if it may yield reasonable results in some situations). 

As a final note, the reader should know that there exists canned finite element analysis software called  NEC4 for computation of fields and energy flow around antennas of relatively arbitrary shape. One option is a plot of energy flow lines in which the local density and direction of lines indicate the magnitude and direction of the flux. Some may prefer that presentation over my Figures 2 and 4. Personally, I rather like the action implied by the arrows in my figures.\\  

\begin{acknowledgments}
I thank Kirk McDonald for drawing my attention to this problem and for stimulating correspondence as well as helpful comments on a draft of this paper.  He oversees a wonderful web site (   http://puhep1.princeton.edu/~mcdonald/examples/EM/   ) with PDF copies of many historic and more recent papers in electromagnetism.  The work was supported in part by the Director, Office of Science, High Energy Physics, U.S. Department of Energy under Contract No. DE-AC03-76SF00098.  
\end{acknowledgments}

\appendix
\section{ SURFACE ELECTRIC FIELD AS EXPANSION IN ASSOCIATED LEGENDRE FUNCTIONS} 

A spherical antenna has a gap in its perfectly conducting surface defined by 
$\ \cos\theta_{1} < \ \cos\theta < \ \cos\theta_{2}$.  The internal source of power creates an electric field $E_{\theta}$ at $r=a$ within the gap, uniformly in azimuth.  Otherwise, $E_{\theta}=0$ on the surface. The multipole expansion (16) of $E_{\theta}$ is in terms of the associated Legendre functions $P_{\ell}^{1}$. We thus require an expansion in those associated Legendre functions of the rectangular function, $f(z) = [\Theta(z-z_{1}) -\Theta(z-z_{2})]$ where $\Theta(x)$ is the Heaviside step function.
We begin with the completeness relation on the interval $(-1, 1)$ in $z=\ \cos\theta$ for the Legendre polynomials $P_{\ell}(z)$ and $P_{\ell}^{1}(z) = -\sqrt{1-z^{2}}dP_{\ell}(z)/dz$:
\begin{eqnarray}
\delta(z-z') & = & \frac{1}{2}\sum_{\ell=0}^{\infty}\ (2\ell+1)\ P_{\ell}(z')P_{\ell}(z)  \label{A1}  \\
\delta(z-z') & = & \frac{1}{2}\sum_{\ell=1}^{\infty}\ \frac{(2\ell+1)}{\ell(\ell+1)}\ P_{\ell}^{1}(z')P_{\ell}^{1}(z)  \label{A2}
\end{eqnarray}
We integrate (\ref{A1}) in \emph{z} over the interval $(-1,z)$ to obtain
\begin{equation}
\Theta(z-z')\ =\ \frac{1}{2}\sum_{\ell=0}^{\infty}\ (2\ell+1)P_{\ell}(z')\ \int_{-1}^{z}P_{\ell}(z)\ dz \label{A3}
\end{equation}
For $\ell=0$ the integral is 
\[\frac{1}{2}\int_{-1}^{z}\ P_{0}(z)\ dz\ =\ (1+z)/2 \] 
and for $\ell>0$,
\[\frac{1}{2}\int_{-1}^{z}P_{\ell}\ dz\ =\ -\frac{1}{2}\int_{z}^{1} P_{\ell}\ dz\ =\ \frac{\sqrt{1-z^{2}}}{2\ell(\ell+1)}P_{\ell}^{1}(z)  \]
The final result can be found in Magnus, Oberhettinger, and Soni.~\cite{MOS}  Substituting these two results into (\ref{A3}) yields
\begin{equation}
\Theta(z-z')\ =\ \frac{1+z}{2} \ + \ \sqrt{1-z^{2}}\sum_{\ell=1}^{\infty}\frac{(2\ell+1)}{2\ell(\ell+1)}P_{\ell}(z')\ P_{\ell}^{1}(z) \label{A4}
\end{equation}

We define the tangential field on the surface as
\begin{equation}
E_{\theta}(r =a,z,z_{1},z_{2})\ =\ \frac{A}{\sqrt{1-z^{2}}}[\ \Theta(z-z_{1})-\Theta(z-z_{2})]   \label{A5}
\end{equation}
where \emph{A} will be chosen for convenience below. With the expansion (\ref{A4}) for $\Theta(z-z')$ the tangential electric field in the gap is given by
\begin{equation}
E_{\theta}(r =a,z,z_{1},z_{2})\ = \ A\ \sum_{\ell=1}^{\infty}\frac{(2\ell+1)}{2\ell(\ell+1)}[\ P_{\ell}(z_{1})-P_{\ell}(z_{2})\ ]\ P_{\ell}^{1}(z)  \label{A6}
\end{equation}

In our calculations we chose the gap to be relatively small and centered around $\ \cos\theta = 0$. With $z_{1}= \ \cos(\pi/2+\epsilon)$ and $z_{2}=\ \cos(\pi/2-\epsilon)$ we find, using (\ref{A5}),  that the voltage \emph{V}, defined as the integral of the electric field across the gap, is $V = 2a\epsilon A$. From the symmetry of the Legendre functions around $\theta = \pi/2$, a symmetric equatorial gap implies only odd $\ell$ terms in (\ref{A6}).  The result for the tangential ($\theta$-component) electric field on the surface of the sphere and in the gap is
\begin{equation}
E_{\theta}(r=a, \ \cos\theta)\ =\ \frac{V}{2a\epsilon}\ \sum_{\ell \  odd}\frac{(2\ell+1)}{\ell(\ell+1)}P_{\ell}(\ \sin\epsilon)\ P_{\ell}^{1}(\ \cos\theta) \label{A7}
\end{equation} 
In the limit of $\epsilon \rightarrow 0$, equation (\ref{A7}) can be shown to be equal to \emph{V/a} times (\ref{A2}) with $z'=0$.


\end {document}